\documentclass[epj]{svjour}
\usepackage{graphics}
\begin{document}
\title{Strongly Coupled Plasma Liquids}
%\subtitle{Do you have a subtitle?\\ If so, write it here}
\author{Z. Donk\'o\inst{1}\thanks{Based on a lecture at the {\it
Zim\'anyi Winter School on Heavy Ion Physics}, Budapest, December 12,
2006. This work has been supported by OTKA-T-48389, OTKA-IN-69892,
MTA/OTKA-90/46140 and OTKA-PD-049991 grants. The authors thank
P. L\'evai for useful discussions on the subject.},
P. Hartmann\inst{1} \and G. J. Kalman\inst{2}} % Do not remove
%
%\offprints{}          % Insert a name or remove this line
%
\institute{Research Institute for Solid State Physics and Optics,
Hungarian Academy of Sciences, POB 49, H-1525 Budapest, Hungary \and
Physics Department, Boston College, Chestnut Hill, MA 02467, USA}
\date{Received: date / Revised version: date}
% The correct dates will be entered by Springer
%
\abstract{ This paper intends to review some of the prominent
properties of strongly coupled classical plasmas having in mind the
possible link with the quark-gluon plasma created in heavy-ion
collisions. Thermodynamic and transport properties of classical
liquid-state one-component plasmas are described and features of
collective excitations are presented.
 % end of PACS codes
} %end of abstract
\maketitle
\section{Introduction}
\label{intro}

In the RHIC experiments Au+Au collisions at ultra-relativistic
energies take place and an extremely high energy density system is
created. The experiments demonstrate that a collective state has been
created in these nucleus-nucleus collisions, where matter consists of
a large number of deconfined quasi-free constituents of the nucleons,
namely quarks and gluons. The striking discovery was that these
particles are, however, in a strongly interacting phase, resembling a
liquid rather than a gas: this phase is being referred to as a
strongly coupled Quark Gluon Plasma (sQGP) \cite{shuryak07}.

The strongly coupled quark-gluon plasma is in many ways similar to
certain kinds of conventional (electromagnetic) plasmas consisting of
electrically charged particles (electrons, ions or large charged
mesoscopic grains), which also exhibit liquid or even solid-like
behavior. These plasmas are known as \emph{strongly coupled plasmas}
and are characterized by an inter-particle potential energy which
dominates over the (thermal) kinetic energy of the particles. Strongly
coupled plasmas occur in electrical discharges, in cryogenic traps and
storage rings, in semiconductors, and in astrophysical systems
(interior of giant planets and white dwarfs). Investigations of these
physical systems have been a major field of activity for some time
\cite{Kalmanbook}.

\begin{figure}
\begin{center}
\resizebox{0.85\columnwidth}{!}{%
\includegraphics{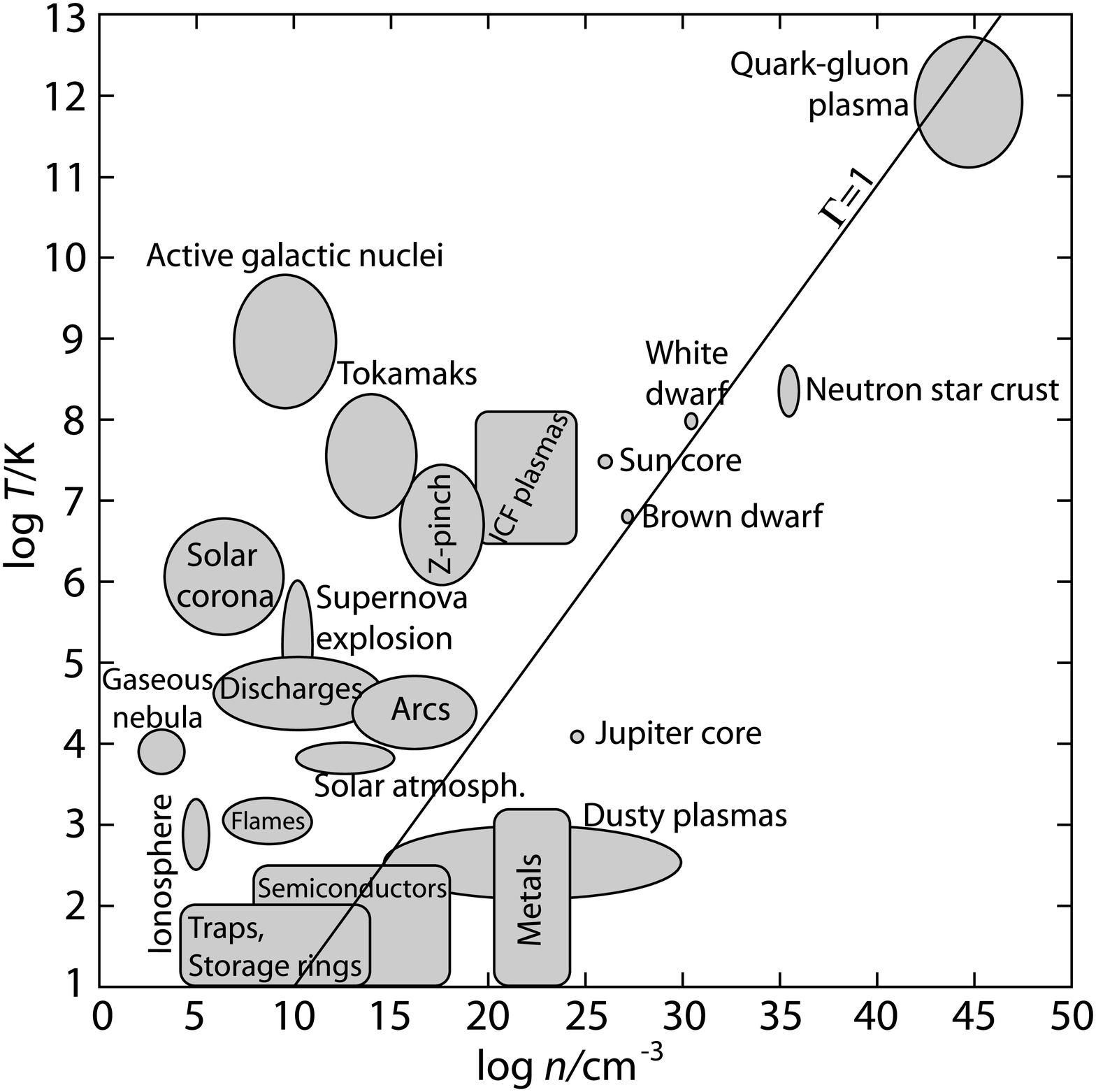}
} \caption{Different types of plasmas over the density --
temperature plane. Note the extremely wide range of these
parameters. Strongly coupled plasmas are located right from the
$\Gamma$ = 1 line. ($\Gamma$ characterizes the ratio of the
potential to kinetic energy, see Eq. (\ref{eq:gamma}).) }
\label{fig:plasmas}
\end{center}
\end{figure}

Plasmas are extremely versatile, as illustrated in
Fig.~\ref{fig:plasmas}, which shows some plasma types over the
density -- temperature plane. Besides the more conventional types of
plasmas, the approximate location of the sQGP is also shown in
Fig.~\ref{fig:plasmas}.

% For one-column wide figures use

Strongly-coupled plasmas may consist of different charged species.
Neutron star crusts are composed of fully stripped iron ions and
electrons. Besides electrons, in the core of Jovian planets we find
a binary mixture of H$^+$ and He$^{2+}$ ions \cite{bim}, while the
core of white dwarf stars consists of a mixture of fully stripped
ions of C, N, and O \cite{cno}. Dusty plasmas, in addition to
electrons and ions, also contain another component of mesoscopic
dust grains, which charge up and respond to electromagnetic fields
similarly as electrons and ions \cite{d1,d2,d3,d4,d5}.

Part of the different plasmas listed above can be described within
the framework of the {\it one-component plasma} model (see e.g.
\cite{BST66},) which considers explicitly only a single type of
charged species and uses a potential that accounts for the presence
and effects of other types of species. This latter may considered as
a charge-neutralizing background, which is either non-polarizable or
polarizable. In the former case the interaction of the main plasma
constituents can be expressed by the
\begin{equation}
\phi(r) = \frac{Q^2}{4 \pi \varepsilon_0} \frac{1}{r} \label{eq:c}
\end{equation}
Coulomb potential energy, whereas in the case of polarizable
background the use of the
\begin{equation}
\phi(r) = \frac{Q^2}{4 \pi \varepsilon_0} \frac{\exp(-r/\lambda_{\rm
D})}{r} \label{eq:y}
\end{equation}
Yukawa potential energy is appropriate to account for screening
effects ($Q$ is the charge of the particles and $\lambda_{\rm D}$ is
the Debye length). As examples of systems for which the Yukawa
potential can be used, dusty plasmas \cite{d1,d2,d3,d4,d5} and
charged colloids \cite{c1a,c1b,c2,c3} may be mentioned.

Strongly coupled plasmas appear in nature and in laboratory
environments in both three-dimensional (3D) and two-dimensional (2D)
settings. While 3D systems are more widespread, notable examples of
2D systems are the layer of dust particles levitated in gaseous
discharges \cite{Thomas94,Chu94} and the layer of electrons over
liquid helium surface \cite{Grimes79,tito}.

Strongly coupled one-component Coulomb systems are fully
characterized by the {\it coupling parameter}:
\begin{equation}\label{eq:gamma}
    \Gamma = \frac{Q^2}{4 \pi \varepsilon_0} \frac{1}{a k_{\rm B} T},
\end{equation}
where $a$ is the Wigner-Seitz (WS) radius, and $T$ is the
temperature. In the case of Yukawa interaction an additional
essential parameter is the {\it screening parameter}:
\begin{equation}\label{eq:kappa}
    \kappa = \frac{a}{\lambda_{\rm D}}.
\end{equation}

The coupling parameter $\Gamma$ is the measure of the ratio of the
average potential energy to the average kinetic energy per particle.
The strong coupling regime corresponds to $\Gamma >$ 1. In the
$\kappa \rightarrow 0$ limit the interaction reduces to Coulomb
type, while at $\kappa \rightarrow \infty$ it approximates the
properties of a hard sphere interaction.

Recent work indicates that the coupling parameter for the sQGP is
expected to be in the order of one \cite{sQGP} (see
Fig.~\ref{fig:plasmas}). This is exactly the reason why methods
traditionally used in the mathematical description of strongly coupled
plasmas may become useful in the physics of sQGP. Therefore, reviewing
the prominent properties of more conventional types of strongly
coupled plasmas is of interest and this is indeed the motivation of
the present paper. We, on the other hand, do not go beyond the
one-component plasma (OCP) model, while a possible description of the
sQGP would clearly require a multicomponent plasma model. The methods
applicable to single-component systems (such as the ones we
deal with in the present paper) serve as the basis of the description
of multicomponent systems, such as ionic mixtures \cite{bim,cno,BIM2}
and charged particle bilayers \cite{bilayer}.

At present, large enough scale numerical simulations, to reproduce
e.g. collective excitations, are only feasible for classical
systems. Therefore we also restrict our studies (presented here) to
classical systems. More sophisticated, but still classical models
aiming partial description of sQGP phenomena including electric and
magnetic charges were developed by the group of Shuryak
\cite{shuryak06}. An attempt to include non-Abelian color-color
interaction into the classical simulation was presented in
\cite{QM05}. It is expected that with the advance of computational
resources large scale simulations for quantum systems will become
realistic within the coming decade.

In Section 2 we introduce the theoretical and numerical methods
applied in our studies. Section 3 describes basic thermodynamic
properties of the strongly coupled one-component plasma (sOCP).
Section 4 deals with the transport properties of sOCP, while Section 5
presents properties of collective excitations characteristic for the
liquid phase sOCP. Finally, Section 6 gives a short summary of the
paper.

\section{Theoretical and numerical methods}

Many body systems can be treated theoretically in a straightforward
way in the extreme limits of both weak interaction and very strong
interaction. In the first case, one is faced with a gaseous system,
or a Vlasov plasma, where correlation effects can be treated
perturbatively ($\Gamma \ll$ 1). Sophisticated theoretical
approaches, like diagrammatic expansions \cite{hansensl} make it
possible to extend standard methods to obtain thermodynamic results
in the moderately coupled regime. The random phase approximation
(RPA) \cite{RPA} method, based on the linear response theory, is a
useful tool to calculate dynamical properties (wave dispersions) in
the case when correlation effects are negligible.

In the case of very strong interaction, the systems crystallize, the
particles are completely localized and phonons are the principal
excitations. For such conditions lattice-summation techniques serve
as solid basis to obtain wave dispersion information.

In the intermediate regime -- in the strongly coupled liquid phase
-- the localization of the particles in the local minima of the
potential surface still prevails, however due to the diffusion of
the particles themselves the time of localization is finite
\cite{caging}. A successful theoretical approach for calculating
structural properties, like the static structure function $S(k)$, is
the Hyper-Netted-Chain (HNC) method with the Percus-Yevik
(PY) equation \cite{HNC}.

The localization of the particles (which may typically cover a period
of several plasma oscillation cycles) serves as the basis of the
\emph{Quasi-Localized Charge Approximation} (QLCA) method
\cite{qlca,qlca2}. The QLCA uses structural information (in the form
of static structure function $S(k)$ or the pair correlation function
$g(r)$) as input for the calculations of the dispersion relations of
the collective modes. The conceptual basis for the QLCA has been a model
that implies the following assumptions about the behavior of strongly
coupled Coulomb or Yukawa liquids \cite{qlca}: (i) in the potential
landscape deep potential minima form that are capable of trapping
(caging) charged particles; (ii) a caged charge oscillates with a
frequency that is determined both by the local potential well and the
interaction with the other (caged) particles in their instantaneously
frozen positions; (iii) the potential landscape changes slowly to
allow the charges to execute a fair number of oscillations; (iv) the
escape from the cages of the particles is caused by the gradual
disintegration of the caging environment; the timescale of this
process is determined by the coupling strength $\Gamma$; (v) the (time
and velocity dependent) correlation between a selected pair of
particles is well approximated by the (time and velocity independent)
equilibrium pair correlation; (vi) the frequency spectrum calculated
from the averaged (correlated) distribution of particles represents,
in a good approximation the average of the distribution of frequencies
originating from the actual ensemble. These assumptions have been
confirmed to be reasonable in a series of studies
(e.g. \cite{caging,qlca2}).

The main concern of the QLCA theory is the analysis of the
collective behavior in strongly coupled many-particle systems. The
formal tools for this are the dielectric function $\varepsilon_{\mu
\nu}^{AB}({\bf k}, \omega)$ [having a tensor character (subscripts)
in real space and a matrix character (superscripts) in species
space] and the dynamical structure function $S^{AB}({\bf k},
\omega)$, or more generally, the dynamical current-current
correlation function $T_{\mu \nu}^{AB}({\bf k}, \omega)$. The
principal approximation of the QLCA method is to replace the
fluctuating microscopic densities and their products by their
ensemble averages, making use of the $S({\bf k})$ static structure
function of the system.

All the information pertaining to the mode structure is contained in
the dielectric matrix that has a longitudinal and a transverse
element:
\begin{equation}
\varepsilon_{\rm {L/T}}(\textbf{k},\omega)=
1-\frac{\omega_0^2(\textbf{k})} {\omega^2-D_{\rm
{L/T}}(\textbf{k})}. \label{eq:eq2}
\end{equation}
Here the $D_{\rm{L}}(\textbf{k})$ and $D_{\rm{T}}(\textbf{k})$ local
field functions are the respective projections of the QLCA dynamical
matrix $D_{\mu \nu}(\textbf{k})$ \cite{qlca}, which is a functional
of the equilibrium pair correlation function (PCF) $h(r) \equiv g(r)
- 1 $ or its Fourier transform $h(\textbf{k})$:
\begin{equation}
D_{\mu \nu}(\textbf{k})= - \frac{n}{m} \int d^2 r M_{\mu \nu}(r)
[e^{i \bf{k} \cdot \bf{r}} -1] h(r) \label{eq:eq3}
\end{equation}
with $ M_{\mu \nu}(r)=\partial_{\mu} \partial_{\nu} \phi(r)$ being
the dipole-dipole interaction potential associated with $\phi(r)$.

The longitudinal and transverse modes are now determined from the
dispersion relations
\begin{equation}
\varepsilon_{\rm{L}}(\textbf{k},\omega)=0, \hspace{1cm}
\varepsilon_{\rm{T}}^{-1}(\textbf{k},\omega)=0. \label{eq:eq4}
\end{equation}

Besides the theoretical approaches computer simulations have proven
to be invaluable tools for investigations of strongly coupled
liquids of charged particles. Monte Carlo (MC) and molecular
dynamics (MD) methods have widely been applied in studies of the
equilibrium and transport properties, as well as of dynamical
effects and collective excitations. The main difference between the
two techniques is that in a MC simulation independent particle
configurations of a canonical ensemble are generated, whereas MD
simulations provide information about the time-dependent phase space
coordinates of the particles, this way allowing studies of dynamical
properties.

Molecular dynamics simulations follow the motion of particles by
integrating their equations of motion while accounting for the
pairwise interaction of the particles, see e.g. \cite{Frenkel}.
Assuming that the dynamics is Newtonian, for each of the particles
the Newtonian equation of motion,
\begin{equation}\label{NMD}
    m \ddot{\bf r}_i = {\bf F}_i
\end{equation}
has to be solved. Here ${\bf F}_i$ is the total force acting on the
$i$-th particle due to all the other particles and due to any
external (e.g. electric and/or magnetic) field. The way how ${\bf
F}_i$ is calculated will be discussed below. In some physical
systems further details have to be considered beyond the Newtonian
approximation. As an example charged colloids can be mentioned where
the Brownian molecular dynamics simulation \cite{c1b,HL92} is widely
used. Brownian molecular dynamics simulations take into account in
the equation of motion solvent friction and a random Langevin force
${\bf R}(t)$ acting on the particles.

In the rest of the paper we restrict our studies to systems where
Newtonian dynamics is a reasonable approximation. In the case of
short-range potentials the calculation of the force acting on a
particle of the system, ${\bf F}_i$, is relatively simple. In this
case MD methods make use of the truncation of the interaction
potential thereby limiting the need for the summation of pairwise
interaction around a test particle to a region of finite size. In the
case of long-range interactions (e.g. Coulomb or low-$\kappa$ Yukawa
potentials), which are of interest here, however, such truncation of
the potential is not allowed, and thus special techniques, like Ewald
summation \cite{Ewald}, have to be used in MD simulations. Besides the
Ewald summation technique there exist few additional methods, like the
fast multipole method and the particle-particle particle-mesh method
(PPPM, or P3M), which can be used to handle long-range interaction
potentials, see e.g.  \cite{Sagui1999}. It is this latter -- widely
used \cite{Hooker,David} -- method, which we also choose as the
simulation approach for our studies presented here. The PPPM method
was originally applied for the case Coulombic interaction
\cite{hockney}. In the PPPM scheme the interparticle force is
partitioned into (i) a force component $F_{\rm PM}$ that can be
calculated on a mesh (the ``mesh force'') and (ii) a short-range
(``correction'') force $F_{\rm PP}$, which is to be applied to closely
separated pairs of particles only. In the mesh part of the calculation
charged clouds are used instead of point-like particles and their
interaction is calculated on a computational mesh, taking also into
account periodic images (for more details see \cite{hockney}). This
way the PPPM method makes it possible to take into account periodic
images of the system (in the PM part), without truncating the long
range Coulomb or low-$\kappa$ Yukawa potentials.  (For high $\kappa$
values the PP part alone provides sufficient accuracy, in these cases
the mesh part of the calculation is not used.)

\section{One-component plasma properties}

Starting with the pioneering work of Brush, Sahlin and Teller
\cite{BST66} and followed by the systematic studies of Hansen and
coworkers \cite{Hansen1,Hansen2,Hansen3,Hansen4} properties of
one-component plasmas have been explored by computer simulation and
theoretical approaches.

In three dimensions (3D) at $\kappa$~=~0 the liquid phase is limited
to coupling parameter values $\Gamma \leq 175$ \cite{Farouki93}. A
first order phase transition was identified to take place at $\Gamma
\cong 175$, where the plasma was found to crystallize into {\it bcc}
lattice \cite{SDS90}. We note that at $\kappa > $~0, the 3D systems 
may crystallize either in {\it bcc} or in {\it fcc} lattices,
depending on the value of $\kappa$ \cite{HFD97}. Crystallization of
the plasma was also experimentally confirmed to take place in many
different systems, e.g. space plasmas \cite{space}, in laser-cooled
trapped ion plasmas \cite{nist} and expanding neutral plasmas
\cite{neutral}, as well as in ion storage rings \cite{beam1,beam2}.

In two dimensions (2D) crystallization into hexagonal lattice occurs
at a lower value of coupling, at $\Gamma \approx 137$, as found by
both computer simulations \cite{Gann79} and by experiments
\cite{Grimes79}. In two dimensions the crystallized form of the
systems is always hexagonal. While most of the studies on 2D system
have been carried out in the crystalline state (``plasma crystals'')
the liquid state also receives more attention nowadays
\cite{NG04,LG05}.

\subsection{Liquid state properties}

\begin{figure}
\begin{center}
\resizebox{0.7\columnwidth}{!}{
\includegraphics{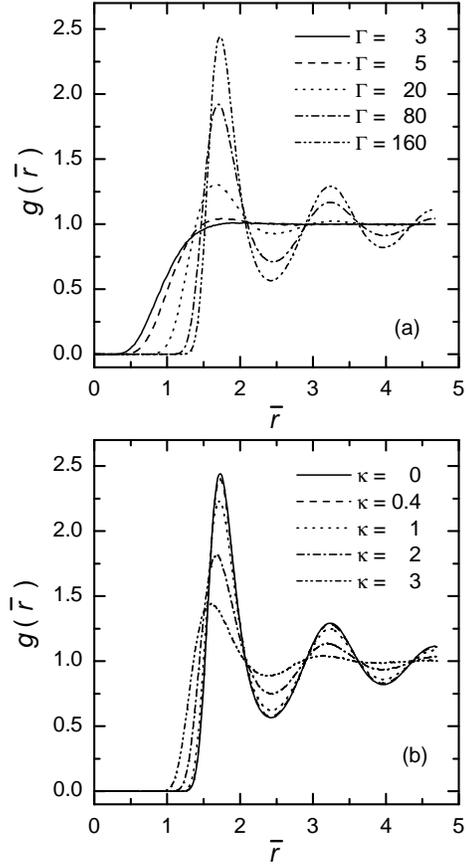}}
\caption{Pair correlation functions of the classical 3D OCP.
Dependence on $\Gamma$ in the Coulomb case (a), and dependence on
the screening parameter of the Yukawa potential at fixed coupling
$\Gamma$ = 160 (b).} \label{fig:pcf}
\end{center}
\end{figure}

At high values of the coupling coefficient plasmas exhibit strong
structural correlations. Such correlations can easily been studied by
examining the pair correlation function
\begin{equation}
g(r)=\frac{V}{4\pi r^2N^2} \left\langle \lim_{{\rm d}r\rightarrow
  0}\frac{1}{{\rm d}r}\sum_i\sum_{j\ne i}\int_r^{r+{\rm
    d}r}\delta(\rho-r_{ij}) ~{\rm d}\rho\right\rangle
\end{equation}
of the system ($V$ is the volume of the system consisting of $N$
particles). Figure~\ref{fig:pcf}(a) shows pair correlation functions
for the Coulomb OCP for a series of coupling parameter values, while
Fig.~\ref{fig:pcf}(b) illustrates the changes of $g(r)$ as a
consequence of screening ($\kappa >$ 0). With decreasing $\Gamma$
the peak amplitudes of the pair correlation function decrease, but
the positions of the peaks remain nearly unchanged. This
remarkable feature of the pair correlation functions indicates that
the local environment of the particles in the liquid phase still
resembles the underlying ($\Gamma \rightarrow \infty$) lattice
configuration. Increasing screening also decreases the degree of
correlation as it can be seen in Fig.~\ref{fig:pcf}(b).

Particle simulation methods provide the full information about the
mechanical state of the system (position and velocity of every
particle in every discrete time step). However, to extract
macroscopic quantities (like thermodynamic energy, pressure,
compressibility) it is useful to construct first distribution
functions (like the $g(r)$ pair correlation function, being
specially important for liquid-state studies, where overall isotropy
is a natural assumption) and use the standard tools of classical
statistical mechanics (e.g. \cite{statphys}) to calculate
thermodynamic characteristics.

The total energy of the system can be expressed as:
\begin{equation}
E=\frac32 Nk_{\rm B}T+U,
\end{equation}
which results in
\begin{equation}\label{eq:energy}
\frac{E}{N}=\frac32 k_{\rm B}T + \frac{n}{2}\int_0^\infty\varphi(r)g(r)~4\pi
r^2 {\rm d}r,
\end{equation}
where $U$ is the excess energy, $n$ is the particle number density
and $\varphi(r)$ is the interaction pair-potential. The pressure can be
calculated using
\begin{equation}
p=nk_{\rm B}T -
\frac{n^2}{6}\int_0^\infty\frac{\partial\varphi(r)}{\partial r}g(r)~4\pi
r^3 {\rm d}r,
\end{equation}
and the isothermal compressibility is expressed by
\begin{equation}
k_{\rm B}T\left(\frac{\partial n}{\partial p}\right)_T =
1+n \int_0^\infty\left[g(r)-1\right]~4\pi r^2 {\rm d}r.
\end{equation}

The application of the above formulae, derived from fundamental
principles of statistical mechanics, depends on the form of the
$\varphi(r)$ pair potential. For Coulomb interaction ($\varphi(r)
\sim 1/r$) the integral in Eq.~(\ref{eq:energy}) is divergent, if the
contribution of the uniform neutralizing background is not properly
taken into account. This can be done simply by replacing $g(r)$ with
$h(r)=g(r)-1$. For Yukawa interaction the contribution of the
polarized background is finite ($E_H\sim 1/\kappa^2$, 3D Hartree
energy). The potential energy can be written as the sum of the
Hartree energy and the correlational energy:
\begin{equation}
\frac{U_{\rm Yukawa}}{N}= \frac{Q^2}{4\pi \varepsilon_0
  ~a}\left[\frac{1}{\kappa^2} +\int_0^\infty h(\bar{r}) \bar{r}
  e^{-\kappa \bar{r}} {\rm d}\bar{r}\right],
\end{equation}
where $\bar{r}=r/a$.

\section{Transport properties}
\label{sec:transport}

In this section we review the data available in the literature for the
basic transport coefficients (self-diffusion, shear viscosity and
thermal conductivity) of the classical one-component plasma. We mainly
present data here for Coulomb OCP, although some data obtained for
Yukawa potential at very low values of the screening parameter ($0.01
\leq \kappa \leq 0.1$) are also shown for comparison. Transport
properties of Yukawa systems characterized by such low $\kappa$ values
are very close to those of Coulomb systems. As already mentioned in
the introduction, there is a growing interest in Yukawa systems. (For
more details about transport properties of Yukawa systems the Reader
is referred to the original publications.)

Simulation techniques have become indispensable tools for the
determination of transport coefficients. The two main approaches of
molecular simulation are the {\it equilibrium} and the {\it
non-equilibrium} methods. In the former one the transport coefficients
are derived from correlation functions of microscopic quantities using
the Green-Kubo (GK) relations. In non-equilibrium simulations a
perturbation is applied to the system and the system's response is
measured.

\subsection{Transport coefficients of the classical one-component plasma}
\label{sec:transport1}

\subsubsection{Self-diffusion}

In equilibrium molecular dynamics simulations one can compute the
self-diffusion coefficient $D$ from either the Green-Kubo relation
\begin{equation}
D = \frac{1}{3} \int_0^\infty \langle {\bf v}(t) {\bf v}(0) \rangle
  {\rm d}t,
\label{eq:diff1}
\end{equation}
(i.e. via the velocity autocorrelation function), or from the
Einstein formula:
\begin{equation}
D = \lim_{t \rightarrow \infty} \frac{1}{6t}
  \langle | {\bf r}_i(t) - {\bf r}_i(0) |^2 \rangle,
\label{eq:diff2}
\end{equation}
(i.e. from the mean square displacement of the particles). In the
above formulae averaging is taken over particles and different
initial times.

\begin{figure}
\begin{center}
\resizebox{0.8\columnwidth}{!}{%
\includegraphics{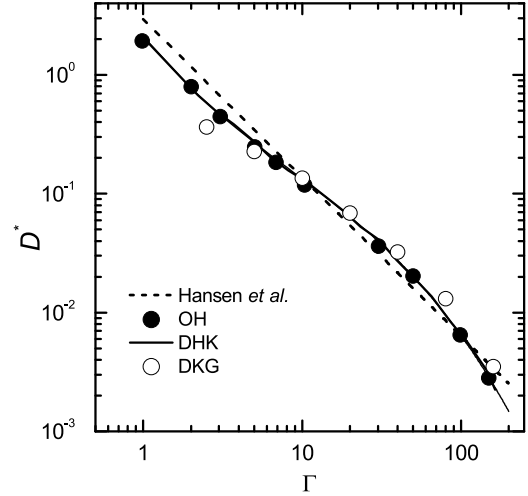}
} \caption{Self-diffusion coefficient of the 3D classical
one-component plasma (OCP). Hansen {\it et al}: \cite{Hansen3}, OH:
Ohta and Hamaguchi \cite{OHdiff}, DKG: Donk\'o, Kalman and Golden,
calculated from cage correlation functions \cite{caging}, DHK:
present MD data. The self-diffusion coefficient has been normalized
as: $D^\ast = D / a^2 \omega_0$. All data correspond to
$\kappa = 0$, except OH, which is for $\kappa = 0.1$.}
\label{fig:ocpdiff}
\end{center}
\end{figure}

The known for the self-diffusion coefficient are displayed in
Fig.~\ref{fig:ocpdiff}. The data have been normalized according to
$D^\ast = D / a^2 \omega_0$, where 
$\omega_0 = \sqrt{n Q^2 / \varepsilon_0 m}$ 
is the plasma frequency and $m$ is the mass of theparticles.

Hansen {\it et al.} made use of (\ref{eq:diff1}) to obtain the
self-diffusion coefficient of the Coulomb OCP. Their results were
found to follow the approximate relation $D^{\ast} = 2.95 \Gamma^{-
1.34}$ \cite{Hansen3}. Ohta and Hamaguchi \cite{OHdiff} obtained the
self-diffusion coefficient for Yukawa liquids from molecular
dynamics simulations using (\ref{eq:diff2}). Their results for
$\kappa$ = 0.1 as well as our present data (based on the same
computational procedure) obtained for $\kappa$ = 0 are also shown in
Fig.~\ref{fig:ocpdiff}. These more recent MD data fall very close to
the those given by the above formula.

An additional set of data derived on the basis of the caged behavior
and jumping of the particles in the strongly coupled liquid phase
\cite{caging} is also shown in Fig.~\ref{fig:ocpdiff}. This data set
agrees surprisingly well with the results of the ``direct'' MD
calculations for $D$ over a wide domain of the coupling parameter.

% For one-column wide figures use

\subsubsection{Shear viscosity}

\begin{figure}
\begin{center}
\resizebox{0.8\columnwidth}{!}{%
\includegraphics{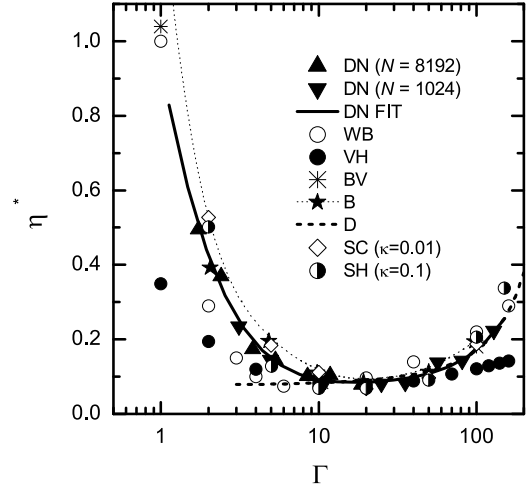}
} \caption{Shear viscosity coefficient of the 3D classical
one-component plasma (OCP). DN: Donk\'o and Ny\'iri \cite{DN00} using
1024 and 8192 particles, WB: Wallenborn and Baus \cite{WB77,WB78},
VH: Vieillefosse and Hansen \cite{Hansen4}, BV : Bernu {\it et al.}
\cite{BVH77,BV78}, B: Bastea \cite{Bastea}, D: Daligault
\cite{Jerome}, SC : Salin and Caillol \cite{SC}, SH : Saigo and
Hamaguchi \cite{SH02}. (The results of \cite{Jerome} have been
scaled to match the minimum value of $\eta$.)} \label{fig:ocpeta}
\end{center}
\end{figure}

The shear viscosity data for the 3D OCP are shown in
Fig.~\ref{fig:ocpeta}. The shear viscosity coefficient has been
normalized as: $\eta^\ast = \eta / m n a^2 \omega_0$.

The shear ($\eta$) and bulk ($\zeta$) viscosity of the 3D OCP was
first derived by Vieillefosse and Hansen \cite{Hansen4} from the
transverse and longitudinal current correlation functions of the
plasma. They have found that the shear viscosity exhibits a minimum at
$\Gamma \approx$ 20. The other main finding of their work was that the
bulk viscosity is orders of magnitude smaller compared to the shear
viscosity. The calculations of Wallenborn and Baus \cite{WB77,WB78}
were based on the kinetic theory of the OCP to calculate $\eta$. Their
results were in a factor of three agreement with the previous results
\cite{Hansen4} at $\Gamma$ = 1 and within a factor of two agreement at
$\Gamma$ = 160. The minimum value of $\eta$ agreed well for both
reports, however the position of the minimum was reported in
\cite{WB77} to occur at a lower coupling value, $\Gamma \approx$
8. Molecular dynamics simulation was first applied by Bernu,
Vieillefosse and Hansen \cite{BVH77,BV78} to obtain transport
parameters through the Green-Kubo relations.  Donk\'o and Ny\'iri
\cite{DN00} used a non-equilibrium MD simulation technique to
determine the shear viscosity, while subsequently, Bastea
\cite{Bastea} applied equilibrium simulation and obtained $\eta$ from
the Green-Kubo relation. Daligault \cite{Jerome} has found that the
shear viscosity of the OCP follows an Arrhenius type behavior at high
$\Gamma$ values. This is shown in Fig.~\ref{fig:ocpeta} by dashed
line.

Salin and Caillol \cite{SC} have carried out equilibrium molecular
dynamics computations for the shear and bulk viscosity coefficients,
as well as for the thermal conductivity of the Yukawa one-component
plasmas. They have implemented Ewald sums for the potentials, the
forces, and for all the currents which enter the Kubo formulas. Saigo
and Hamaguchi \cite{SH02} have also used the Green-Kubo relations for
the calculations of the shear viscosity. As a refinement, the effect
of the plasma environment is dusty plasmas has been taken into account
through Langevin dynamics in the calculation of shear viscosity of
Yukawa systems \cite{KAZ}.

% For one-column wide figures use

\subsubsection{Thermal conductivity}

% For one-column wide figures use
\begin{figure}
\begin{center}
\resizebox{0.8\columnwidth}{!}{%
\includegraphics{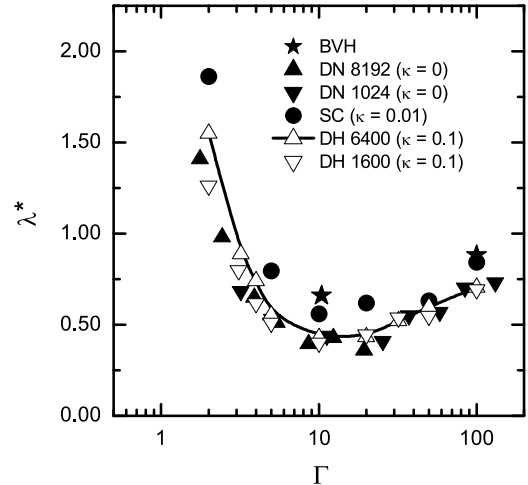}
} \caption{Thermal conductivity coefficient of the 3D classical
one-component plasma (OCP). BVH: Bernu {\it et al.}
\cite{BVH77,BV78},  DN: Donk\'o and Ny\'iri \cite{DN00} using
1024 and 8192 particles, SC: Salin and Caillol \cite{SC}, 
DH: Donk\'o and Hartmann \cite{DH04}.}
\label{fig:ocplambda}
\end{center}
\end{figure}

Thermal conductivity data for the 3D OCP are shown in
Fig.~\ref{fig:ocplambda}. The thermal conductivity coefficient has
been normalized as: $\lambda^\ast = \lambda / n k_{\rm B} a^2
\omega_0$.

We have reported non-equilibrium molecular dynamics calculation of the
thermal conductivity of the classical OCP in \cite{DNSH98}. In
contrast with the studies of Bernu {\it et al} \cite{BVH77,BV78} where
the transport coefficients were obtained from the simulation of an
equilibrium system, we applied a perturbation to the system and
deduced $\lambda^{\ast}$ from the relaxation time of the system
towards the equilibrium state. Donk\'o and Hartmann \cite{DH04}
applied the non-equilibrium MD method proposed by M\"uller-Plathe
\cite{FMPheat} to calculate the thermal conductivity of Yukawa
liquids. 

As regards to Yukawa systems, transport parameters have been studied
in several papers. Besides the work of Salin and Caillol \cite{SC}
(mentioned above), Faussurier and Murillo obtained thermal
conductivity (as well as self-diffusion and shear viscosity) values
for the Yukawa OCP through its mapping with the Coulomb OCP system,
based on the Gibbs-Bogolyubov inequality \cite{FM}.

\section{Collective behavior}

Collective excitations (waves) are prominent features of
plasmas. Depending on the dimensionality and the confinement of the
system, different collective excitations (longitudinal and transverse
modes) may show up. Longitudinal modes can fully be characterized by
the dynamical structure function $S(k,\omega)$, while transverse modes
can be studied through the analysis of the transverse current
fluctuation spectra $T(k,\omega)$. The corresponding current
fluctuation spectra for the longitudinal mode, $L(k,\omega)$, is
linked with the dynamical structure function through $L(k,\omega) =
(\omega^2 / k^2) S(k,\omega)$. Collective excitations are identified
as peaks in these spectra and dispersion relation are derived by
observing the change of the frequency (where the peaks are found) with
wave number. Additionally, the widths of the peaks in the spectra
convey information about the lifetime of excitations (associated with
the damping of the waves), as well as about the distribution of the
mode frequencies due to the disordered particle configuration in the
liquid phase.

Collective effects in Coulomb \cite{KG90,GKW92} and Yukawa
\cite{OH2000,HO2000,RosKal97} plasmas have extensively been
investigated. In Coulomb systems the longitudinal (plasmon) mode is
known to have a frequency $\omega \rightarrow \omega_{\rm p}$ at $k
\rightarrow 0$. The transverse mode exhibits an acoustic dispersion.
Rosenberg and Kalman \cite{RosKal97} investigated the dispersion
relation for dust acoustic waves in a strongly coupled dusty plasma
with the aid of the QLCA scheme, generalized to take into account
electron and/or ion screening of the dust grains. Hamaguchi and Ohta
\cite{OH2000,HO2000} studied the wave dispersion relations in the
fluid phase of Yukawa systems through molecular dynamics
simulations. They have demonstrated that the transverse wave
dispersion has a cutoff at a long wavelength even in the case of weak
screening. Their results have confirmed the earlier theoretical
predictions \cite{RosKal97}. The QLCA method has subsequently been
applied to determine the properties of the transverse (shear) mode in
strongly coupled dusty plasmas \cite{KRDW00}. For this mode the
dispersion was found be characterized by a low-$k$ acoustic behavior
and by a frequency maximum lying well below the plasma frequency.  The
dispersion curves of the longitudinal and transverse modes were
demonstrated to merge at high wave number around the Einstein
frequency of localized oscillations. Experimental observation of
transverse shear waves in the strongly coupled liquid phase of a
three-dimensional (layered) dusty plasma have been reported by
Pramanik {\it et al.} \cite{Kaw}. The collective modes of dusty
plasmas in the liquid phase have also been investigated theoretically
by Murillo \cite{Murillo1,Murillo2,Murillo3}.

In the following we present MD simulation results for the collective
excitations in 3D Yukawa liquids, and compare these with the
predictions of the QLCA theory. The simulations have been carried
out using $N$ = 12800 particles. In the MD simulation information
about the (thermally excited) collective modes and their dispersion
is obtained from the Fourier analysis of the correlation spectra of
the density fluctuations
\begin{equation}\label{eq:rho}
\rho(k,t)= k \sum_{j=1}^{N} \exp \bigl[ i k x_j(t) \bigr]
\end{equation}
yielding the dynamical structure function as \cite{Hansen3}:
\begin{equation}\label{eq:sp1}
S(k,\omega) = \frac{1}{2 \pi N} \lim_{\Delta T \rightarrow \infty}
\frac{1}{\Delta T} | \rho(k,\omega) |^2,
\end{equation}
where $\Delta T$ is the length of data recording period and
$\rho(k,\omega) = {\cal{F}} \bigl[ \rho(k,t) \bigr]$ is the Fourier
transform of (\ref{eq:rho}).

% For one-column wide figures use
\begin{figure}
\begin{center}
\resizebox{0.7\columnwidth}{!}{% 
\includegraphics{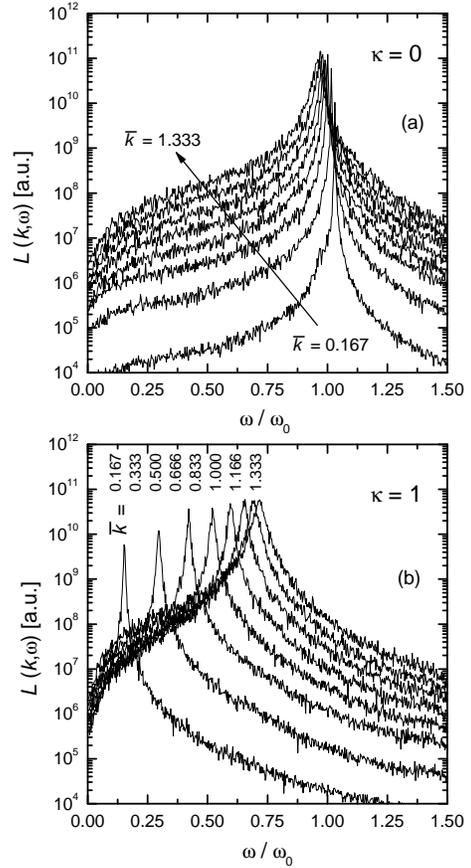} 
} \caption{Longitudinal current fluctuation spectra in 3D Coulomb
OCP at $\Gamma$ = 160 and in a Yukawa OCP at
$\Gamma$ = 200, $\kappa$ = 1. $\bar{k} = k a$ denotes the
dimensionless wave number, its values are given in (b), while in (a)
the arrow indicates increasing values of $\bar{k}$.}
\label{fig:spectra-L}
\end{center}
\end{figure}
%

% For one-column wide figures use
\begin{figure}
\begin{center}
\resizebox{0.7\columnwidth}{!}{% 
\includegraphics{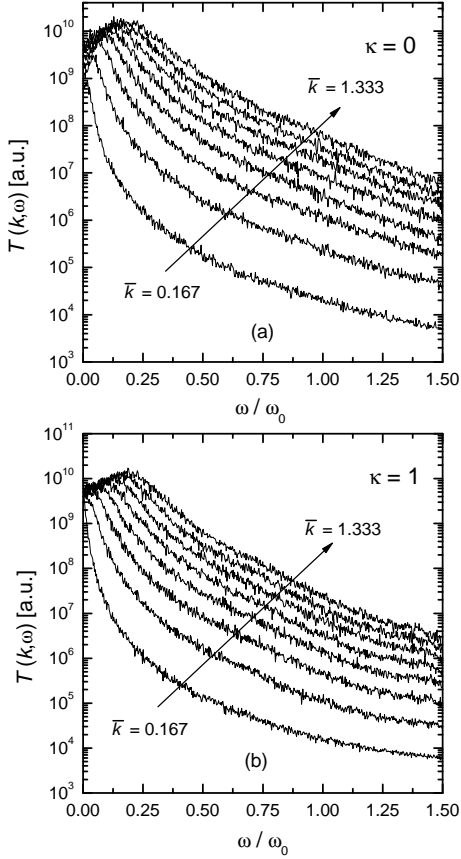} 
} \caption{Transverse current fluctuation spectra in 3D Coulomb
OCP at $\Gamma$ = 160 and in a Yukawa OCP at
$\Gamma$ = 200, $\kappa$ = 1. The arrows indicates increasing values
of the dimensionless wave number $\bar{k}$ (for values see
Fig.~\ref{fig:spectra-L}).} \label{fig:spectra-T}
\end{center}
\end{figure}

Similarly, the spectra of the longitudinal and transverse current
fluctuations, $L(k,\omega)$ and $T(k,\omega)$, respectively, can be
obtained from Fourier analysis of the microscopic quantities
\begin{eqnarray}
\lambda(k,t)&=& k \sum_{j=1}^{N} v_{j x}(t) \exp \bigl[ i k x_j(t) \bigr], \nonumber \\
\tau(k,t)&=& k \sum_{j=1}^{N} v_{j y}(t) \exp \bigl[ i k x_j(t) \bigr],
\label{eq:dyn}
\end{eqnarray}
where $x_j$ and $v_j$ are the position and velocity of the $j$-th
particle. Here we assume that ${\bf k}$ is directed along the $x$
axis (the system is isotropic) and accordingly omit the vector
notation of the wave number. The way described above for the
derivation of the spectra provides information for a series of wave
numbers, which are multiples of $k_{\rm min} = 2 \pi / H$, where $H$
is the edge length of the simulation box. The collective modes are
identified as peaks in the fluctuation spectra. The widths of the
peaks provide additional information about the lifetimes of the
excitations: narrow peaks correspond to longer lifetimes, while
broad features are signals for short lived excitations.

\begin{figure}
\begin{center}
\resizebox{0.8\columnwidth}{!}{%
\includegraphics{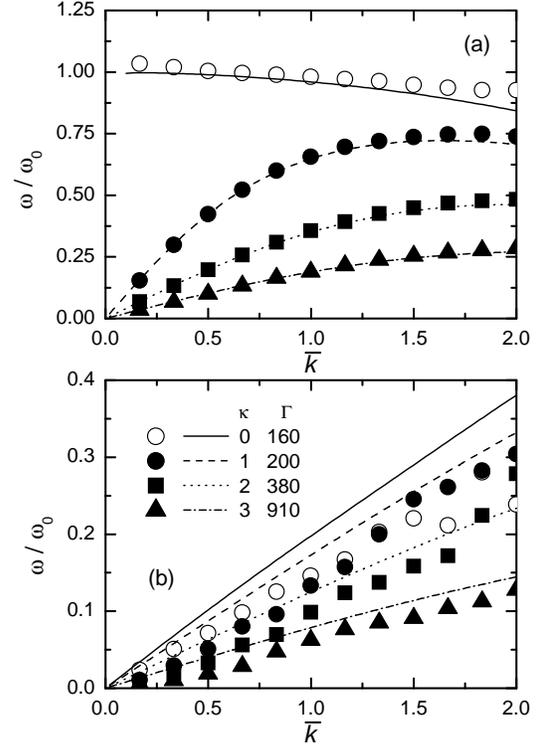}
} \caption{Dispersion relations for the (a) longitudinal and (b)
transverse modes of 3D Coulomb and Yukawa plasma liquids. Symbols
represent molecular dynamics results, while the lines correspond to
the predictions of the QLCA theory. The ($\Gamma,\kappa$) pairs are
given in the legend of panel (b).} \label{fig:disp}
\end{center}
\end{figure}

Representative longitudinal and transverse current fluctuation
spectra, $L(k,\omega)$ and $T(k,\omega)$, respectively, are plotted
in Figs.~\ref{fig:spectra-L} and \ref{fig:spectra-T} for wave
numbers, which are multiples of $\overline{k }_{\rm min}= {k}_{\rm min} a$
= 0.167. $L(k,\omega)$ obtained for the Coulomb case ($\Gamma$ =
160, $\kappa$ = 0) peaks very nearly at the plasma frequency
$\omega_0$. In the presence of screening (Yukawa potential),
as shown in Fig.~\ref{fig:spectra-L}(b), the behavior of
$L(k,\omega)$ changes significantly: at $\bar{k} \rightarrow$ 0 the
wave frequency $\omega \rightarrow$ 0. The contrast between the
$\kappa$ = 0 and the $\kappa >$ 0 cases is also well seen in
Fig.~\ref{fig:disp}, where the dispersion curves derived from the
fluctuation spectra are displayed. The dispersion curves for $\kappa > 0$ 
are quasi-acoustic ($\omega/\omega_0 \propto \bar{k}^{1/2}$),
with a linear portion near $k$ = 0, which gradually extends when
$\kappa$ is increased. The ($\Gamma$,$\kappa$) pairs for which the
dispersion graphs are plotted in Fig.~\ref{fig:disp} have been
selected to represent a constant ``effective'' coupling
$\Gamma^\ast$ = 160. This definition of $\Gamma^\ast$ relies on the
constancy of the first peak amplitude of the pair correlation
function $g(\bar{r})$, similarly to the case of 2D Yukawa liquids
\cite{H2005}.

Compared to those characterizing the $\cal{L}$ mode, peaks in the
$\cal{T}$ mode spectra are rather broad, as it can be seen in
Fig.~\ref{fig:spectra-T}(a) and (b), for the Coulomb and Yukawa
cases, respectively. In the case of this mode there is no significant
change between the behavior when $\kappa$ changes from zero to a
nonzero value, only the mode frequency decreases, as can be observed
in Fig.~\ref{fig:disp}(b).

% For one-column wide figures use

Comparison of the dispersion relations obtained from the MD and QLCA
results \cite{KRDW00} is presented in Fig.~\ref{fig:disp}. The QLCA
equations for the mode frequencies need the pair correlation
function as input data. The data shown in Fig.~\ref{fig:disp} were
obtained using MD-generated $g(r)$ functions. The agreement between
the (MD and QLCA) dispersion curves is excellent for the $\cal{L}$
mode, while some differences in the frequency of the $\cal{T}$ waves
can be seen in Fig.~\ref{fig:disp}(b). This latter may originate
from the inaccurate determination of the peak positions of the
rather spread $T(k,\omega)$ spectra. Another difference is the
cutoff of the $\cal{T}$ mode dispersion curve at finite wave
numbers. This disappearance of the shear modes for $k \rightarrow$ 0
is a well known feature of the liquid state
\cite{Hansen3,TK80,SZRT97} while the sharp cut-off $\omega
\rightarrow$ 0 for a finite $k$ has also been observed in
simulations of Yukawa systems \cite{OH2000,Murillo3}. It is noted
that this cutoff is not accounted for by the QLCA, as it does not
include damping effects.

\section{Summary}

This paper intended to review some of the important properties of
strongly coupled plasmas (within the framework of the one-component
plasma (OCP) model), which might be relevant to the studies of
strongly interacting quark-gluon plasma (sQGP). 

At high values of the coupling coefficient ($\Gamma \gg 1$) the
one-component plasma exhibits liquid state properties. In this domain
the pronounced peaks of the pair correlation function indicate the
presence of strong correlation effects. The transport coefficients --
self-diffusion, shear viscosity and thermal conductivity -- have
thoroughly been investigated and their behavior is quite well
understood. While the self-diffusion coefficient decreases
monotonically with increasing $\Gamma$, the shear viscosity and
thermal conductivity exhibit a minimum at moderate values of coupling
($10 \leq \Gamma \leq 20$), due to the different temperature
dependence of the kinetic and potential contributions to these
transport coefficients. Two types of collective excitations -- a
longitudinal and a transverse mode -- have been identified in the OCP
system, the emergence of latter of them being attributed to strong
correlations.

\end{document}